\documentclass{ws-procs9x6}

\begin{document}

\title{ Tetraquark states and spectrum.}

\author{Elena Santopinto  and Giuseppe Galat\`a}

\address{ INFN and Universit\`a di Genova,
via Dodecaneso 33, 16146 Genova, Italy}

\begin{abstract}

A general classification of tetraquark states in terms of the spin-flavour, colour and spatial 
degrees of freedom has been constructed.
The permutational symmetry properties of both the spin-flavour and orbital parts of the
quark-quark and antiquark-antiquark subsystems are discussed in short.
This classification is model indipendent and useful both for model-builders and experimentalists.
An evaluation of the tetraquark spectrum is obtained from a generalization of a Iachello mass formula, originally developed for the $q\bar q$ mesons. The ground state tetraquark nonet is identified with
$f_{0}(600)$, $\kappa(800)$, $f_{0}(980)$, $a_{0}(980)$.  
\end{abstract}

\bodymatter

\section{Introduction}\label{sec:introduzione}

A new interest in light meson spectroscopy has been triggered recently by the KLOE, E791 and BES collaborations, which have provided evidence of the low mass resonances $f_{0}(600)$ \cite{Aloisio:2002bt,Aitala:2000xu,Ablikim:2005ni} and $\kappa(800)$ \cite{Aitala:2000xu,Ablikim:2005ni}. Maiani et al. \cite{Maiani:2004uc} have suggested that the lowest lying scalar mesons, $f_{0}(980)$, $a_{0}(980)$, $\kappa(800)$ and 
$f_0(600)$ could be described as a tetraquark nonet, in particular as a diquark and antidiquark system, since the quark-antiquark assignment to P-waves \cite{Tornqvist:1995kr} has never
 worked in the scalar case \cite{Jaffe:1976ig}. 
In the traditional quark-antiquark scheme, the $f_{0}(980)$ is associated with non-strange quarks \cite{Tornqvist:1995kr} and so it is difficult to explain both its higher mass respect to the other components of the nonet and its decay properties \cite{Jaffe:1976ig,Maiani:2004uc}.  Already in the seventies Jaffe \cite{Jaffe:1976ig} suggested the tetraquark structure of the scalar nonet and proposed a four quark bag model. Other identifications have been proposed (for a complete review see \cite{Amsler:2004ps,Close:2002zu,PDG} and references therein), in particular 
as quasimolecular-states \cite{Weinstein:1982gc} and as dynamically generated resonances \cite{Oset}.

We construct a complete classification scheme of the two quark-two antiquark states in terms of SU(6)$_{~sf}$ \cite{Santopinto:2006my}.
An evaluation of the tetraquark spectrum for the lowest scalar meson nonet is obtained \cite{Santopinto:2006my} from a generalization, to the tetraquark case, of the Iachello mass formula
for normal mesons \cite{Iachello:1991fj}.

\section{The classification of tetraquark states}\label{sec:classificazionestati}

The tetraquark wave function contains contributions connected to the spatial and the internal (colour, flavour and spin) degrees of freedom.
We shall make use of symmetry principles without, for the moment, introducing any explicit dynamical model.
In the construction of the classification scheme we are constrained by two conditions: the tetraquark wave functions should be a colour singlet, as all physical states, and the tetraquarks states must be antisymmetric for the exchange of the two quarks and the two antiquarks.

The allowed SU(3)$_{f}$ representations for the $qq\bar q\bar q$ system are obtained by means of the product $[3]\otimes [ 3]\otimes [\bar 3]\otimes [\bar 3] =[1]\oplus [8]\oplus [1]\oplus [8]\oplus [27]\oplus [8]\oplus [8]\oplus [10]\oplus [\overline{10}]$. The allowed isospin values are $I=0,\frac{1}{2},1,\frac{3}{2},2$, while the hypercharge values are $Y=0,\pm 1,\pm 2$. The values $I=\frac{3}{2},2$ and $Y=\pm 2$ are exotic, which means that they are forbidden for the $q\bar q$ mesons.
The allowed SU(2)$_{s}$ representations are obtained by means of the product $ [2]\otimes [2]\otimes [2]\otimes [2] =[1]\oplus [3]\oplus [1]\oplus [3]\oplus [3]\oplus [5]$. 
The tetraquarks can have an exotic spin $S=2$, value forbidden for normal $q\bar q$ mesons.
The SU(6)$_{sf}$-spin-flavour classification is obtained
 by $[6]\otimes [ 6]\otimes [\bar 6]\otimes [\bar 6]=[1]\oplus [35]\oplus [405]\oplus [1]\oplus [35]\oplus [189]\oplus [35]\oplus [280]\oplus [35]\oplus [\overline{280}]$. In Appendices A and B of Ref.~\refcite{Santopinto:2006my} all the flavour and spin states in the $qq\bar q \bar q$ configuration are explicitly written. Moreover in the ideal mixing hypothesys, the flavour states of the tetraquarks are a superposition of the SU(3)-symmetrical states in such a way to have defined strange quark and antiquark numbers. Clearly the only states that can be mixed are those with the same good quantum numbers, i.e. same isospin and hypercharge, and they are explicitly constructed in Ref.~\refcite{Santopinto:2006my}. The ideal mixing is essentially a consequence of the OZI rule and it is a hypothesys that has to be proved yet, but it is used by all the authors  working on $q\bar q$ and tetraquarks.
In a system made up o four objects, like the tetraquark, we have to define three relative coordinates that we choose as in Ref.~\refcite{Estrada}: two relative coordinates between two quarks and two antiquarks and a relative coordinate between their centers of mass to which are associated three orbital angular momenta, $L_{13}$, $L_{24}$ and $L_{12-34}$ respectively.  
We have four different spins and three orbital angular momenta and a total angular momentum $J$ obtained by combining spins and orbital momenta.
The parity for a tetraquark system is the product of the intrinsic parities of the quarks (+) and the antiquarks (-) times the factors coming from the spherical harmonics  \cite{Estrada} as
$P=P_{q}P_{q}P_{\bar q}P_{\bar q}(-1)^{L_{13}}(-1)^{L_{24}}(-1)^{L_{12-34}}=(-1)^{L_{13}+L_{24}+L_{12-34}}$.
Only the states for which $Q$ and $\bar Q$, where $Q$ represents the couple of quarks and $\bar Q$ the couple of antiquarks, have opposite charges are $C$ eigenvectors, with eigenvalues \cite{Estrada} $C=(-1)^{L_{12-34}+S}$. The eigenvalues of the G parity are in Ref.~\refcite{Santopinto:2006my}.
Tetraquark mesons do not have forbidden $J^{PC}$ combinations. 
The tetraquark states must be antisymmetric for the exchange of the two quarks and the two antiquarks and so it is necessary to study the permutational symmetry (i.e. the irreducible representations of the group $S_{2}$) of the colour, flavour, spin and spatial parts of the wave functions of each subsystem. Moreover we have another constraint:   
only the singlet colour states are physical states. Regarding the two colour singlets allowed to the tetraquarks, it is better to write them by underlining their permutational $S_{2}$ symmetry, antisymmetric (A) or symmetric (S): $(qq)$ in $\;[\bar 3]_{C}\;(A)\;$ and $\;(\bar q\bar q)\;$ in $ [3]_{C}\; (A)$, or  $(qq)\;$ in $ [6]_{C} (S)$ and $\; (\bar q\bar q)\;$ in $\; [\bar 6]_{C}\; (S)$. 
The following step is studying the permutational symmetry of the spatial part of the two quarks (two antiquarks) states and the permutational symmetry of the SU(6)$_{sf}$ representations for a couple of quarks. The spatial, flavour, colour and spin parts with given permutational symmetry ($S_{2}$) must then be combined together to obtain completely antisymmetric states under the exchange of the two quarks and the two antiquarks. The resulting states are listed in Table 
III of  Ref.~\refcite{Santopinto:2006my}.
In Table V, VI, VII and VIII of Ref.~\refcite{Santopinto:2006my} the possible flavour, spin and $J^{~PC}$ values for different orbital angular momenta are studied. 

In 1991 Iachello, Mukhopadhyay and Zhang developed a mass formula \cite{Iachello:1991fj} for $q\bar q$ mesons,  
\begin{equation}
M^{2}=(N_{n}M_{n}+N_{s}M_{s})^{2}+a \, \nu +b\, L+c\, S+d\, J+e\, M'^{2}_{iji'j'}+f\, M''^{2}_{iji'j'},
\label{eq:formulamassa}
\end{equation}
where $N_{n}$ is the non-strange quark and antiquark number, $M_{n}\equiv M_{u}=M_{d}$ is the non-strange constituent quark mass, $N_{s}$ is the strange quark and antiquark number, $M_{s}$ is the strange constituent quark mass, $\nu $ is the vibrational quantum number, $L$, $S$ and $J$ are the total orbital angular momentum, the total spin and the total angular momentum respectively, $M'^{2}_{iji'j'}$ and $M''^{2}_{iji'j'}$ are two phenomenological terms which act only on the lowest pseudoscalar mesons. The first acts only on the octect and encodes the unusually low masses of the eight Goldstone bosons, while the second acts on the $\eta $ and $\eta '$ mesons and encodes the non-negligible $q\bar q$ annihilation effecs that arise when the lowest mesons are flavour diagonal.
The flavour states are considered in the ideal mixing hypothesis, with the exception of the lowest pseudoscalar nonet.
During the 15 years that have passed from the publication of Iachello's article the values reported by the PDG regarding the mesons are changed in a considerable way, so we have decided to updated the fit of the Iachello model using the latest values reported by the PDG \cite{PDG} for the light $q \bar q$ mesons . The resulting parameters are reported in Ref.~\refcite{Santopinto:2006my}. 
\begin{table}
\tbl{The candidate tetraquark nonet. Experimental data and quantum numbers}
{\begin{tabular}{@{}lllll@{}}\toprule
Meson & $I^{G}(J^{PC})$ & $ N_s$ & $Mass ~(GeV)$ & Source \\\colrule
$a_{0}(980)$ & $1^{-}(0^{++})$ & 2 & $0.9847\pm 0.0012$  & PDG \cite{PDG} \\
$f_{0}(980)$ & $0^{+}(0^{++})$ & 2 & $0.980\pm 0.010$ &  PDG \cite{PDG}  \\
$f_{0}(600)$ & $0^{+}(0^{++})$ & 0 & $0.478\pm 0.024$ &  KLOE \cite{Aloisio:2002bt}  \\
$k(800)$ & $\frac{1}{2}(0^{+})$ & 1 & $0.797\pm 0.019$ &  E791 \cite{Aitala:2002kr}  \\\botrule
\end{tabular}}
\label{tab:nonet}
\end{table}
The Iachello mass formula was developed for $q\bar q$ mesons. In order to describe uncorrelated tetraquark systems by means of an algebraic model one should use a new spectrum generating algebra for the spatial part, i.e. U(10). We have not addressed this difficult problem, but we chose to write the internal degrees of freedom part of the mass formula in the same way as it was done for the $q\bar q$ mesons. The splitting inside a given flavour multiplet, to which is also associated a given spin, can be described by the part of the mass formula that depends on the numbers of strange and non-strange quarks and antiquarks. 
Thus we can use, with the only purpose of determining the mass splitting of the candidate tetraquark nonet, see Ref.~\refcite{Santopinto:2006my},
\begin{equation}
\label{eq:tetrascorrelati}
M^{2}=\alpha +(N_{n}M_{n}+N_{s}M_{s})^{2},
\end{equation}
where $\alpha $ is a constant that includes all the spatial and spin dependence of the mass formula, and $M_{n}$ and $M_{s}$ are the masses of the constituent quarks as obtained from the previously discussed upgrade of the parameters of the Iachello mass formula.
We set the energy scale, i. e. we determine the constant $\alpha $, by applying Eq. (\ref{eq:tetrascorrelati}) to the best-known candidate tetraquark, $a_{0}(980)$, see Ref.~\refcite{Santopinto:2006my}. Thus, the masses of the other mesons belonging to the same tetraquark nonet, predicted with our simple formula, are $M(\kappa(800))=0.726\;GeV$, $M(f_{0}(600))=0.354\;GeV$ and $M(f_{0}(980))=0.984\;GeV$ .
These values seem in agreement with the experimental values, even if, before reaching any conclusion, new experiments are mandatory.

\end{document}